\documentclass[aps,twocolumn,groupedaddress,showpacs]{revtex4-1}
\usepackage{graphicx}
\usepackage{dcolumn} 
\usepackage{bm}
\usepackage{amsbsy}
\usepackage{amssymb}
\usepackage{arrayjobx}
\usepackage{amsmath,amsfonts}
\usepackage{mathtools} 
\usepackage{multirow}
\usepackage{braket}

\begin{document}

\title{Excitations in the quantum paramagnetic phase of the quasi-one-dimensional
Ising magnet CoNb$_2$O$_6$ in a transverse field: Geometric
frustration and quantum renormalization effects}

\author{I.~Cabrera}
\affiliation{Clarendon Laboratory, University of Oxford, Parks
Road, Oxford OX1 3PU, United Kingdom}
\author{J.~D.~Thompson}
\affiliation{Clarendon Laboratory, University of Oxford, Parks
Road, Oxford OX1 3PU, United Kingdom}
\author{R.~Coldea}
\affiliation{Clarendon Laboratory, University of Oxford, Parks
Road, Oxford OX1 3PU, United Kingdom}
\author{D.~Prabhakaran}
\affiliation{Clarendon Laboratory, University of Oxford, Parks
Road, Oxford OX1 3PU, United Kingdom}
\author{R.~I.~Bewley}
\affiliation{ISIS Facility, Rutherford Appleton Laboratory,
Chilton, Didcot, Oxon OX11 0QX, United Kingdom}
\author{T.~Guidi}
\affiliation{ISIS Facility, Rutherford Appleton Laboratory,
Chilton, Didcot, Oxon OX11 0QX, United Kingdom}
\author{J.~A. Rodriguez-Rivera}
\affiliation{NIST Center for Neutron Research, National Institute
of Standards and Technology, Gaithersburg, Maryland 20899, USA}
\author{C.~Stock}
\affiliation{NIST Center for Neutron Research, National Institute
of Standards and Technology, Gaithersburg, Maryland 20899, USA}

%\date{\today}

\begin{abstract}
The quasi-one-dimensional (1D) Ising ferromagnet CoNb$_2$O$_6$ has
recently been driven via applied transverse magnetic fields
through a continuous quantum phase transition from spontaneous
magnetic order to a quantum paramagnet, and dramatic changes were
observed in the spin dynamics, characteristic of weakly perturbed
1D Ising quantum criticality.  We report here extensive
single-crystal inelastic neutron scattering measurements of the
magnetic excitations throughout the three-dimensional (3D)
Brillouin zone in the quantum paramagnetic phase just above the
critical field to characterize the effects of the finite
interchain couplings. In this phase, we observe that excitations
have a sharp, resolution-limited line shape at low energies and
over most of the dispersion bandwidth, as expected for spin-flip
quasiparticles. We map the full bandwidth along the
strongly dispersive chain direction and resolve clear modulations
of the dispersions in the plane normal to the chains,
characteristic of frustrated interchain couplings in an
antiferromagnetic isosceles triangular lattice. The dispersions
can be well parametrized using a linear spin-wave model that
includes interchain couplings and further neighbor exchanges. The
observed dispersion bandwidth along the chain direction is smaller
than that predicted by a linear spin-wave model using exchange
values determined at zero field, and this effect is attributed to
quantum renormalization of the dispersion beyond the spin-wave
approximation in fields slightly above the critical field, where
quantum fluctuations are still significant.
\end{abstract}


\maketitle

\section{Introduction}

\label{sec:introduction} Quantum phase transitions are
characterized by a qualitative change in the ground state of a
system that occurs upon varying an external parameter at zero
temperature. The one-dimensional (1D) quantum Ising chain in
transverse magnetic field is one of the most theoretically studied
paradigms for a continuous quantum phase transition [\onlinecite{Sachdev}].
Here, a transverse magnetic field promotes quantum fluctuations in
a ground state where spins are initially spontaneously
ferromagnetically aligned ``up" or ``down" along an Ising axis.
When these fluctuations are strong enough (compared to mean-field
effects), the spontaneous Ising order is suppressed giving way to a
quantum paramagnetic phase, where spins are in a correlated
superposition of ``up" and ``down" states. Although this model was
solved exactly more than four decades ago (via mapping to
Jordan-Wigner fermions [\onlinecite{Pfeuty}]), an experimental realization
of this theoretical paradigm was only recently
achieved [\onlinecite{Coldea}] in the quasi 1D Ising-like ferromagnet
CoNb$_2$O$_6$. Experiments observed that at a critical field
applied transverse to the Ising axis, a phase transition occurred
from the spontaneously ordered state, characterized by 1D domain
wall (kink) excitations, into the quantum paramagnetic phase,
characterized by sharp, spin-flip quasiparticles. Moreover, in the
ordered phase, a rich spectrum of two-kink bound states was
observed (and even more structure was recently resolved by THz
spectroscopy [\onlinecite{THz}]), and understood in terms of confinement
effects due to an effective longitudinal mean field resulting from
the interchain couplings. Near the critical field, the ratios of
the energies of the two lowest bound states approached the golden
ratio, in agreement with long-standing, field-theory predictions
for a universal E8 spectrum for the critical Ising chain perturbed
by a weak longitudinal field [\onlinecite{zamolodchikov}].

In understanding the rich physics of CoNb$_2$O$_6$, the presence of
the weak interchain couplings is essential, as they are responsible
for the mean-field effects that lead to two-kink bound states via
confinement effects in the ordered phase. Recent theoretical
work [\onlinecite{lee2010}] has highlighted potentially even richer physics
due to the fact that the interchain couplings form a distorted
(isosceles) triangular lattice with antiferromagnetic couplings.
The resulting frustration effects, combined with the strong quantum
fluctuations tuned by the transverse field, have been predicted to
stabilize a fine structure inside the ``ordered'' part of the
phase diagram. As many as four distinct ordered phases are predicted
(including a zero-temperature incommensurate spin-density-wave
state stabilized exclusively by quantum fluctuations) depending on
the level of the isosceles distortion from the perfect
(equilateral) triangular lattice. It is therefore of fundamental
interest to obtain direct information about the geometry and
strength of the interchain couplings and quantify the degree of
the isosceles distortion of the triangular lattice.

The most direct measurement of the interchain couplings is via the
dispersion of the excitations in the plane perpendicular to the
magnetic chains. In the ordered phase, where excitations are
two-kink bound states, the interchain dispersion is much
suppressed [\onlinecite{Carr}] and only occurs to higher order in the
strength of the interchain exchanges. The situation is very
different in the quantum paramagnetic phase above the critical
field, where the spontaneous order has disappeared. In this phase,
the excitations can be understood as a first approximation by
starting from the high-field limit, where they are
coherently-propagating single spin flips that can hop along all
links of the lattice to first order in the corresponding exchange
coupling strengths. Therefore, part of the motivation behind the
experiments reported here is to probe with high-resolution
inelastic neutron scattering the full 3D dispersion of the
excitations in the quantum paramagnetic phase above the critical
field and extract quantitatively the strength of the interchain
couplings.

The phase above the critical field is described as a quantum
paramagnet [\onlinecite{Sachdev}]. It is paramagnetic in the sense that the
magnetization along the field is not yet saturated in the region
immediately above the critical field (for the 1D Ising chain at
the critical field, only about {\em half} the moment is polarized
along the field [\onlinecite{Pfeuty}]), such that there is a large part of
the magnetic moment available that, in principle, could
spontaneously order due to the Ising exchange. However, coherent
quantum fluctuations induced by the transverse field suppress such
order (as opposed to a thermal paramagnet at high temperature,
where the spontaneous magnetic order is suppressed by random
thermal fluctuations). The quantum paramagnet is smoothly
connected (without any phase transition, only a cross-over) to the
fully-polarized phase reached in the asymptotic limit of very high
fields. The excitations in this whole region of the phase diagram
can be understood by starting from the high-field limit, where they
are coherently-propagating single spin flips with a large (Zeeman)
gap. Upon decreasing the field towards the critical field, quantum
fluctuations increase and subsequently they decrease the
magnetization along the field. At low energies, sharp, well-defined
quasi-particles are expected throughout this phase, protected from
decay because of the finite gap. However, the fundamental quantum
nature of those quasiparticles is far from trivial.  They
originate from single spin flips, but are ``dressed" by strong
quantum fluctuations, and their dispersion relation may also be
strongly renormalized compared to a semi-classical spin-wave
description, which assumes that excitations are {\em literally}
plane-wave superpositions of single spin flips. Therefore, another
objective of the experiments reported here is to probe
experimentally the full bandwidth of the dominant, along-chain
dispersion in the quantum paramagnetic phase, to see if the
excitations are sharp over the full energy scale of the spectrum
and test to what extent the dispersion can be quantitatively
described by a spin-wave approach or whether quantum
renormalization effects are relevant.
%%%%%%%%%%%%%%%%%%%%%%%%%%%%%%%%%%%%
\setcounter{figure}{0}
\begin{figure}[t!]
\renewcommand{\thefigure}{1}
\includegraphics[width=\linewidth,clip=true, trim=20 40 15 0] {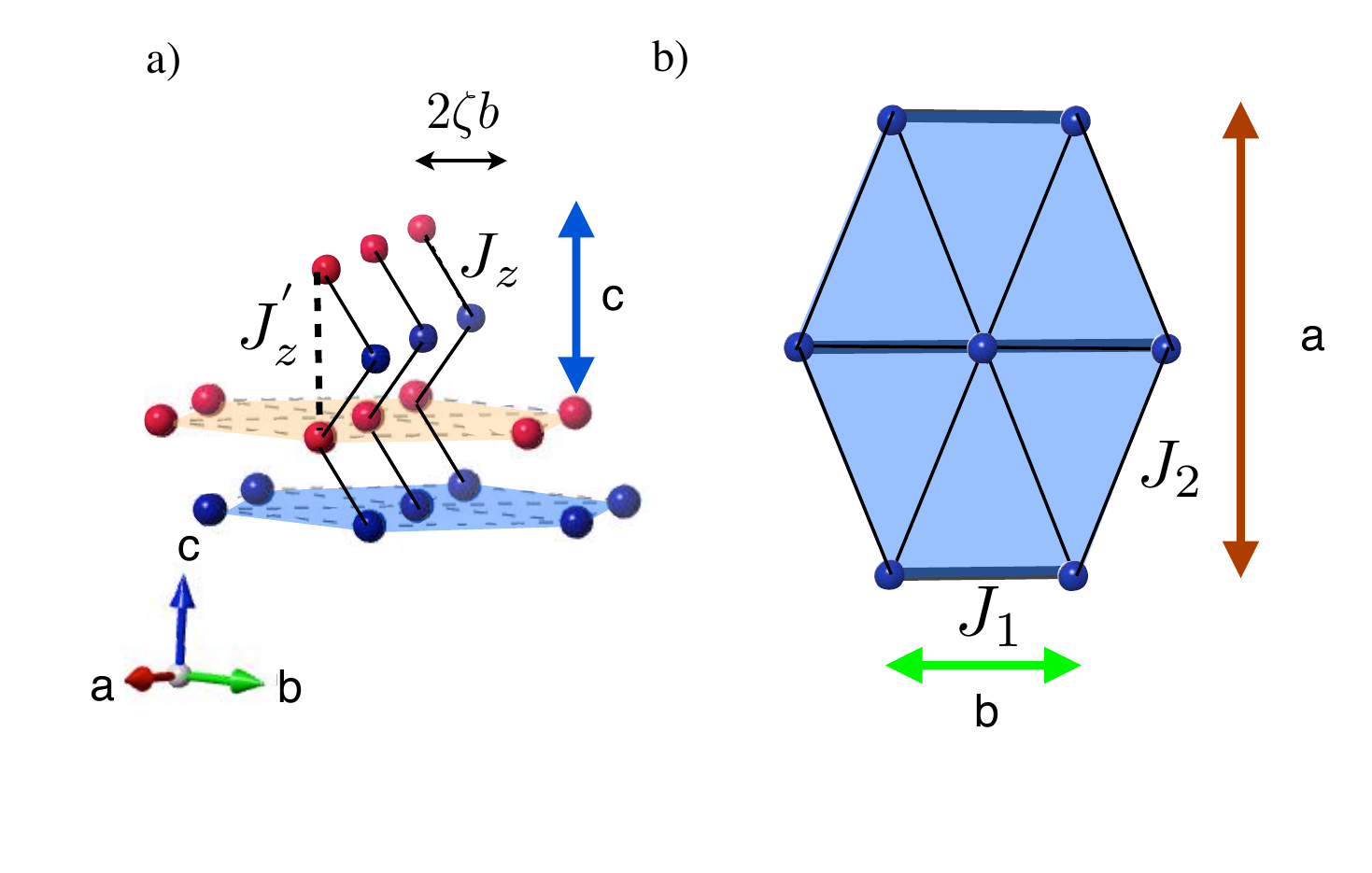}
\caption{\label{fig:structure} (Color online) Lattice of Co$^{2+}$
ions and relevant (Ising) exchange paths. (a) Zigzag chains
running along $c$ (and buckled along $b$) have a dominant
ferromagnetic nearest-neighbor coupling $J_z$ and weaker
next-nearest-neighbor exchange $J'_z$. (b) The buckled magnetic
chains form two inequivalent isosceles triangular lattices in the
basal $ab$ plane with interchain interactions $J_1$ (along the
$\pm\bm{b}$ bonds) and $J_2$ (along the $(\pm\bm{a}\pm\bm{b})/2$
bonds). For clarity, only the blue triangular lattice is shown.}
\end{figure}
%%%%%%%%%%%%%%%%%%%%%%%%%%%%%%%%%%%%%%

The crystal structure of CoNb$_2$O$_6$ is orthorhombic (space
group $Pbcn$) and the magnetic ions are Co$^{2+}$ occupying a
single crystallographic site [\onlinecite{Heid1995123}] [4$c$
$(0,\zeta,1/4)$ with $\zeta=0.165$] in a lattice of zigzag
magnetic chains along the crystallographic $c$ axis, with a
triangular lattice arrangement in the basal $ab$ plane, as
illustrated in Fig.~\ref{fig:structure}. Due to a combination of
strong crystal-field and spin-orbit coupling effects the magnetic
ground state of the Co$^{2+}$ ($3d^7$) ions is a Kramers doublet
(effective spin $S=1/2$) with a magnetic moment with a strong
preference to point along a local easy axis (Ising direction $z$),
located in the $ac$ plane at a finite angle ($\gamma =
29.6^{\circ}$) to the $c$ axis [\onlinecite{Heid1995123}]. The magnetic
interactions between neighboring Co$^{2+}$ moments have been
proposed to be of the Ising form $S^z_i S^z_j$ with the strongest
interaction a ferromagnetic coupling $J_z$ between
nearest-neighbors along the chain, followed by weaker
antiferromagnetic (AFM) couplings $J'_z$ between
next-nearest-neighbors along the chain, and much weaker interchain
couplings $J_1$ and $J_2$ along the bonds of the isosceles
triangular lattice in the $ab$ plane, both
AFM [\onlinecite{Coldea,Heid1995123}], see Fig.~\ref{fig:structure}.

In zero applied magnetic field, the finite interchain couplings
stabilize magnetic order below 2.95\, K in a structure where spins
are ordered ferromagnetically along the magnetic chains and the
ordered spin magnitude varies between chains following an
incommensurate wave vector $(0,q,0)$ with $q=0.37$ just below the
transition temperature [\onlinecite{Heid1995123,Heid1997574}]. Such an
incommensurate spin-density-wave order is the natural ordering
instability in an isosceles Ising triangular lattice, where at the
onset temperature
$q=\frac{1}{\pi}\cos^{-1}\left(\frac{J_2}{2J_1}\right)$ (see
Refs.~[\onlinecite{Heid1995123, Bak1980}]). The ordering wave vector
was observed to be temperature-dependent upon cooling and to
lock-in to the commensurate value $(0,1/2,0)$ at $1.97$\, K, below
which the order is antiferromagnetic with a constant-magnitude
ordered spin on every site. For magnetic fields applied along the
$b$ axis [\onlinecite{Coldea}], transverse to the Ising axes, the
spontaneous magnetic order is entirely suppressed at $5.5$\, T, and
it is above this field that all measurements reported here have
been collected.

The rest of the paper is organized as follows. Section
\ref{sec:experimental} gives details of inelastic neutron
scattering experiments performed to probe the magnetic excitations
in the quantum paramagnetic phase in a high transverse field, the
results of those experiments are presented in
Sec.~\ref{sec:results}. The observed dispersion relations are
parameterized first in Sec.~\ref{sec:hopping} in terms of a
phenomenological model of (spin-flip) quasiparticles that
propagate by (spin-isotropic) hopping terms between the sites of
the magnetic lattice. This captures well the low-energy
modulations of the dispersion and the overall dispersion shape
given the lattice topology of chains with a triangular lattice
geometry in the basal plane. The model also accounts for the
presence of an additional weaker intensity shadow mode, attributed
to the magnetic unit cell doubling induced by the buckling of the
magnetic chains. This model however does not capture the observed
suppression of the interchain dispersion at high energies. In
Sec.~\ref{sec:lswt} we compare the magnetic excitations with
predictions of a microscopic spin exchange Hamiltonian with
dominant Ising coupling along the chain direction and additional
further neighbor exchanges solved in the linear spin-wave
approximation. We first present a one-sublattice approximation in
Sec.~\ref{sec:lswt_one} and then a four-sublattice model
appropriate for the orthorhombic crystal structure in
Sec.~\ref{sec:lswt_four}. We show that the latter can provide a
very good parametrization of the observed dispersions, intensity
dependence in the Brillouin zone and relative intensity between
the main and shadow modes. In Sec.~\ref{sec:discussion} we discuss
the fact that the observed dispersion bandwidth is smaller than
the calculated bandwidth using a linear spin-wave approach using
exchange values estimated earlier from a parametrization of the
spin dynamics in zero applied field. We propose that the observed
smaller dispersion bandwidth is due to a quantum renormalization
of the dispersion at fields not too high above the critical
transverse field not captured by spin-wave theory. We discuss this
effect in detail for the pure Ising chain in a transverse field by
comparing the known exact quantum solution with linear spin-wave
results. Finally, conclusions are summarized in
Sec.~\ref{sec:conclusions}.

\section{Experimental Details}
\label{sec:experimental} A 7\, g single crystal of CoNb$_2$O$_6$
grown using the floating zone technique [\onlinecite{Prabhakaran}] was
mechanically fixed inside a custom-made oxygen-free copper can to
prevent sample movement due to strong torques that arise when an
external magnetic field is applied transverse to the Ising axis of
the spins at low temperatures. The crystal was aligned in the
horizontal $(h0l)$ plane, which contains both the $c$ direction of
the magnetic chains and the Ising direction. Throughout this
paper, wave-vector components $(h,k,l)$ are given with reference to
the reciprocal lattice of the  crystallographic orthorhombic unit
cell with lattice parameters $a=14.1337$ \AA, $b=5.7019$ \AA,
$c=5.0382$ \AA~ at 2.5\, K from Ref.~[\onlinecite{Heid1995123}]. The
sample mount was attached to the bottom of a dilution refrigerator
insert with a base temperature of 0.03\, K. Magnetic fields were
applied along the $b$ axis (transverse to the Ising axes of all
spins) using a vertical superconducting magnet. All measurements
reported here were made in fields between 7 and 9\, T in the
quantum paramagnetic phase above the quantum critical phase
transition at 5.5\, T.

Inelastic neutron scattering experiments were performed using both
the multiangle triple-axis MACS [\onlinecite{MACS}] spectrometer at the
NIST Center for Neutron Research as well as the direct-geometry
time-of-flight chopper spectrometer LET [\onlinecite{LET}] at ISIS. MACS
was operated to measure the inelastic scattering of neutrons with
a fixed final energy $E_f=3$\, meV as a function of wave-vector
transfer in the horizontal $(h0l)$ scattering plane at constant
energy transfer (from $E=0.4$ to 2.75\, meV). This enabled probing
the magnetic excitations along the chain direction $l$, as well as
along the interchain direction $h$, with typical counting times of
2 h to collect a complete wave-vector map at a fixed energy
transfer.

On LET, the inelastic scattering was probed for neutrons with
incident energies of $E_i=2.1, 4$, and $10$\, meV with a measured
energy resolution (FWHM) on the elastic line of 0.023(1), 0.051(1),
and 0.21(1)\, meV, respectively. LET was operated to record the
time-of-flight data for incident neutron pulses of all the above
different energies simultaneously. The large vertical opening of
the magnet on LET allowed probing the inelastic scattering using
position sensitive detectors along the direction perpendicular to
the scattering plane, i.e., the $k$ direction, which was essential in
order to obtain a complete map of the dispersions in the full 3D
Brillouin zone. The inelastic scattering was measured for a
selection of fixed sample orientations to probe the full bandwidth
of the magnetic dispersion along the chain direction $l$ and also
along the interchain $h$ and $k$ directions near the along-chain
ferromagnetic zone center $l=0$, with typical counting times of 2
h per fixed sample orientation setting. Furthermore, a series
of measurements at fixed sample orientations spanning an angular
range of 90$^\circ$ in $1^\circ$ steps (counting time 3 mins/step)
were combined into a Horace [\onlinecite{horace}] four-dimensional volume file
to extract the complete magnetic dispersion maps in the $(hk0)$
plane perpendicular to the magnetic chains. All data sets were
collected at fixed temperatures between 0.03 and 1.8\, K. Even the
highest temperature satisfied the criterion that it was much
smaller than the spin gap (0.389\, meV at 7\, T), so thermal effects
to the measured dispersions are expected to be negligible.

%%%%%%%%%%%%%%%%%%%%%%%%%%%%%%%%
\begin{figure}[tbh]
\includegraphics[width=\linewidth,clip=true,trim= 5 0 5 0] {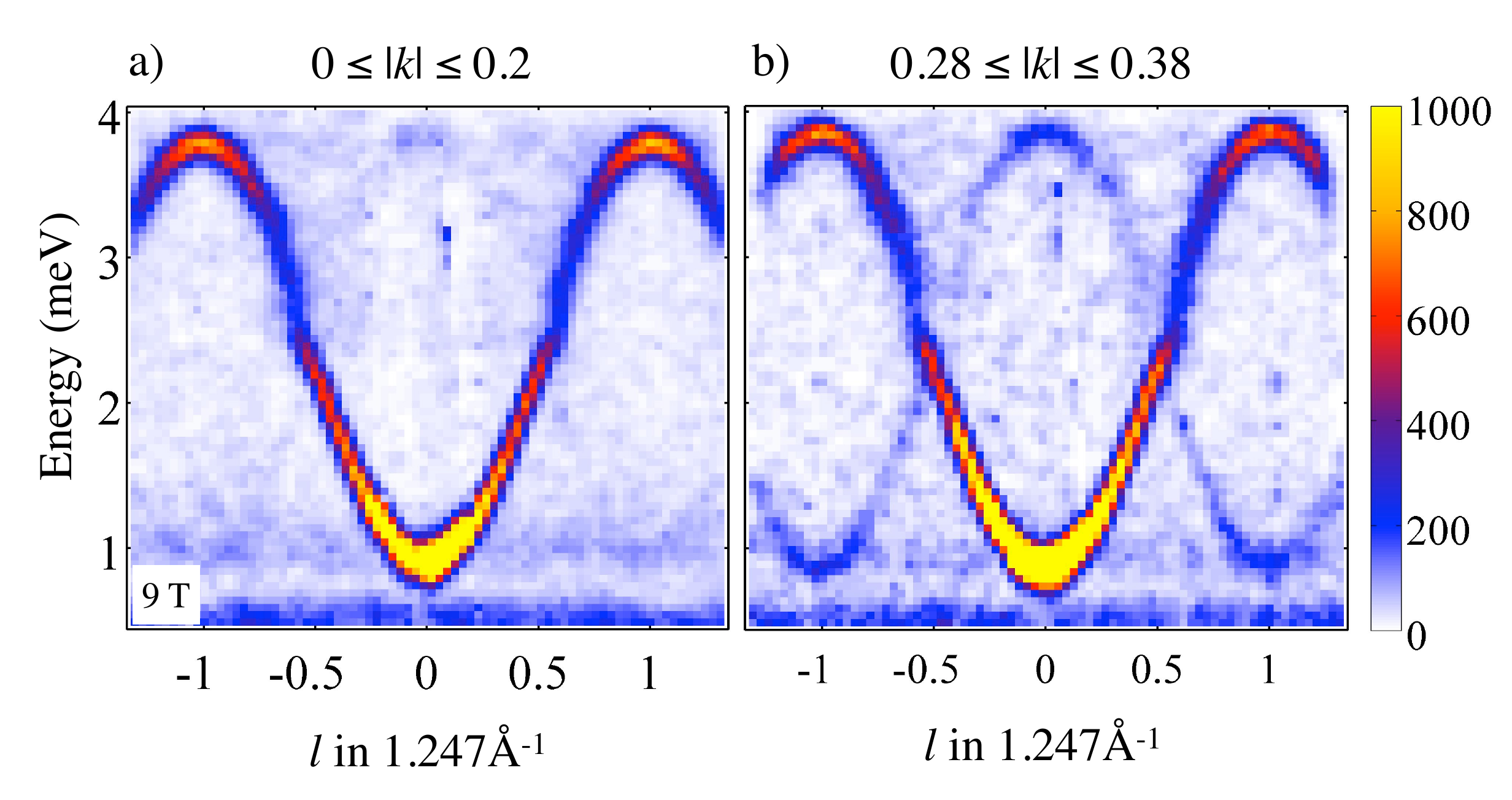}
\caption{\label{fig:Ldisp10meVall} (Color online) Dispersion along
the chain direction $l$ in the paramagnetic phase at 9\, T and
$\sim$0.03\, K, measured on LET with an incident energy
$E_i=10$\, meV. (a) A single dispersive mode is observed for small
interchain wave-vector component $k$ and (b) a second,
weaker-intensity mode becomes visible for finite $k$. The plots
show the averaged neutron scattering intensity for $|k|$ in the
range [0.0,0.2] in (a) and [0.28,0.38] in (b).}
\end{figure}
%%%%%%%%%%%%%%%%%%%%%%%%%%%%%%%%%%%%%%

\section{Measurements and results}
\label{sec:results}
\subsection{Dispersion along the chain direction}
The observed inelastic neutron scattering spectrum at 9\,T shown
in Fig.~\ref{fig:Ldisp10meVall}(a) is dominated by a single mode
with a small linewidth at low energies and over most of the
dispersion bandwidth. The largest dispersion bandwidth is along the chain
direction (2.85\,meV) with a near-sinusoidal shape with the
minimum (0.92\, meV) at the zone center $l=0$ and periodicity $l
\rightarrow l+2$, as expected for dominant ferromagnetic coupling
between spins spaced by $c/2$ along the chain direction, see
Fig.~\ref{fig:structure}(a). For finite interchain wave-vector
component $k$, a much weaker intensity, ``shadow'' mode is also
observed with the same dispersion relation as the main mode, but
shifted by $l \rightarrow l+1$, see
Fig.~\ref{fig:Ldisp10meVall}(b). As we will show later, this is due
to the fact that the magnetic chains are not straight, but buckled
[see Fig.~\ref{fig:structure}(a)], this buckling leads to an
effective doubling of the magnetic unit cell along $c$ (compared
to straight chains) and zone folding leads to the observed shadow
mode.
%%%%%%%%%%%%%%%%%%%%%%%%%%%%%%%%%%%%%%
\begin{figure*}[htb]
\includegraphics[width=\linewidth, clip=true,trim=0 0 0 0] {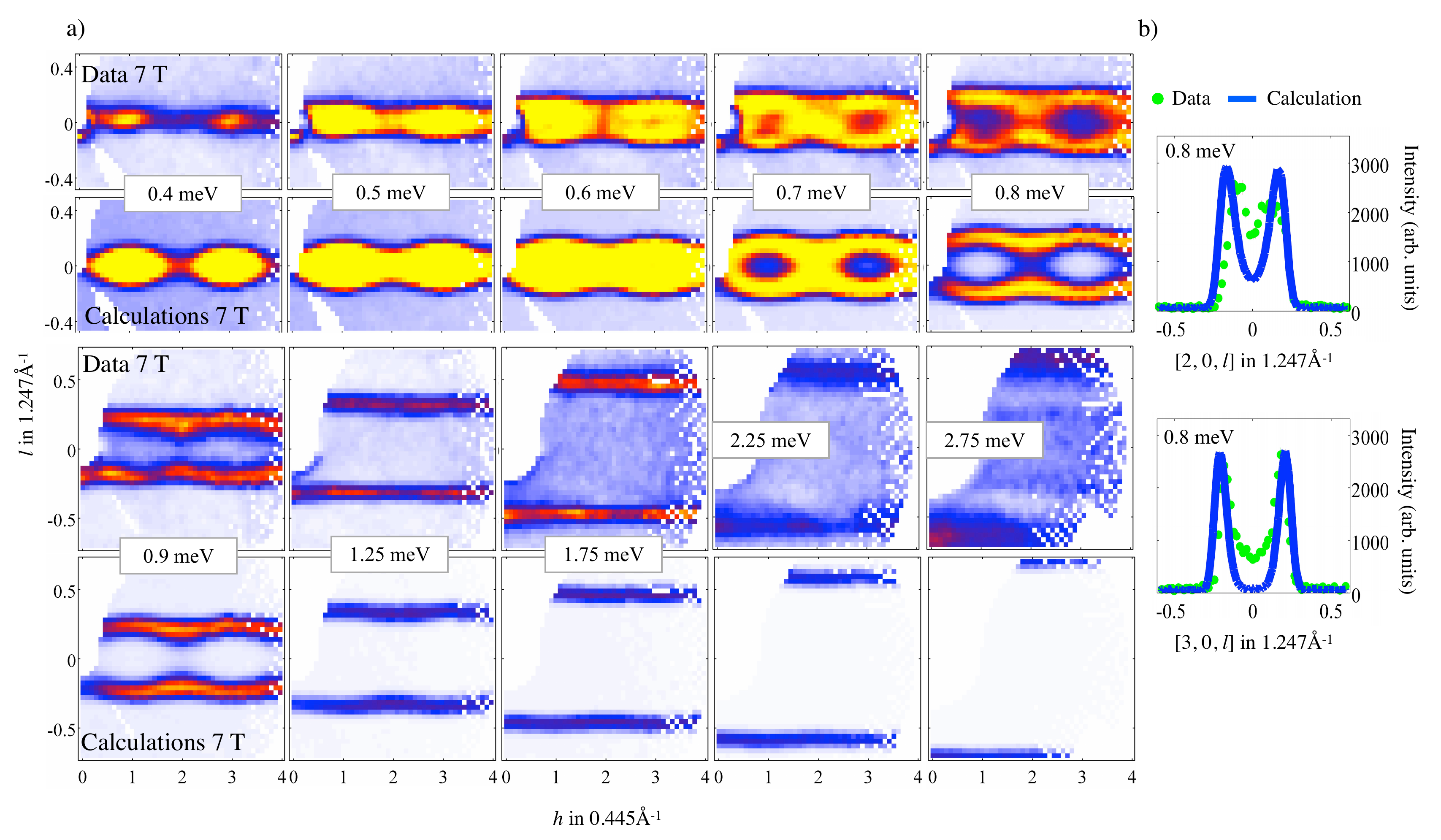}
\caption{\label{fig:macs0p40p9} (Color online) (a) Inelastic
neutron scattering intensity at constant energy transfer $E$ as a
function of wave vector in the ($h0l$) plane measured at 7\, T and
0.5\, K on MACS. Rows 1 and 3 are data, rows 2 and 4 are
calculations using the one-sublattice spin-wave model in
Eq.~(\ref{hw}) with intensities given in Eq.~(\ref{intensities})
and an overall single intensity scaling factor for all the panels.
The calculations include the neutron polarization factor, the
spherical magnetic form factor for Co$^{2+}$, and instrumental
resolution effects. At low energies, near the dispersion minimum,
strong dispersion is seen along both $h$ and $l$. As the energy
increases, the dispersion along the interchain direction $h$
becomes flatter, suggesting that interchain couplings become less
relevant at higher energies. (b) Wave-vector scans along (2,0,$l$)
and (3,0,$l$) at constant energy transfer $E=0.8$~meV showing
displacement of the spin-wave peaks upon changing $h$, data (green
circles) and spin-wave model (blue line). Error bars represent
$\pm 1$ standard deviation.}
\end{figure*}
%%%%%%%%%%%%%%%%%%%%%%%%%%%%%%%%%%%%%%%%

\subsection{Interchain dispersion}
To probe the sensitivity of the dispersion relations to the
interchain couplings, detailed measurements of the inelastic
spectrum were first performed in the ($h0l$) plane at a somewhat
lower field of 7\, T (still in the quantum paramagnetic phase) and
are shown in Fig.~\ref{fig:macs0p40p9}. Constant energy maps of
the inelastic neutron scattering are plotted along the interchain
direction $h$ (horizontally) and along the chain direction $l$
(vertically) for energy transfers $E$ starting from the minimum of
the dispersion near 0.4\, meV and up to 2.75\, meV. Compared to the
9~\, T data in Fig.~\ref{fig:Ldisp10meVall}(a) at this lower field
of 7\, T the spin gap is reduced due to the decrease in Zeeman
energy. The regions of strong intensity indicate the location of
the constant energy contours in the dispersion surface. For
decoupled magnetic chains along $c$, the dispersion would be
expected to depend only on $l$ and be independent of $h$, so
constant energy contours would be expected to be parallel
horizontal lines that move further apart in $l$ with increasing
energy. The data, however, shows very clear modulations of the
constant energy contours along both $h$ and $l$, in particular at
the lower energies towards the minimum gap. For example, a
dispersion along $h$ is very clearly seen at $E=0.8$\, meV in
Fig.~\ref{fig:macs0p40p9}(a)(top right panel) and it is also very
pronounced at the lowest energies $E=0.4$\, meV (top left panel)
where minima in the shape of rugby-balls are centered at odd $h$
positions. The data indicate that interchain couplings produce
important modulations in the dispersion relation at low energies
near the minimum gap with weaker effects (less interchain
dispersion) at higher energies above $\sim$$1.25$\, meV.

%%%%%%%%%%%%%%%%%%%%%%%%%%%%%%%%%%%%

\begin{figure*}[p]
\includegraphics[width=\linewidth,clip=true,trim= -35 0 -40  0] {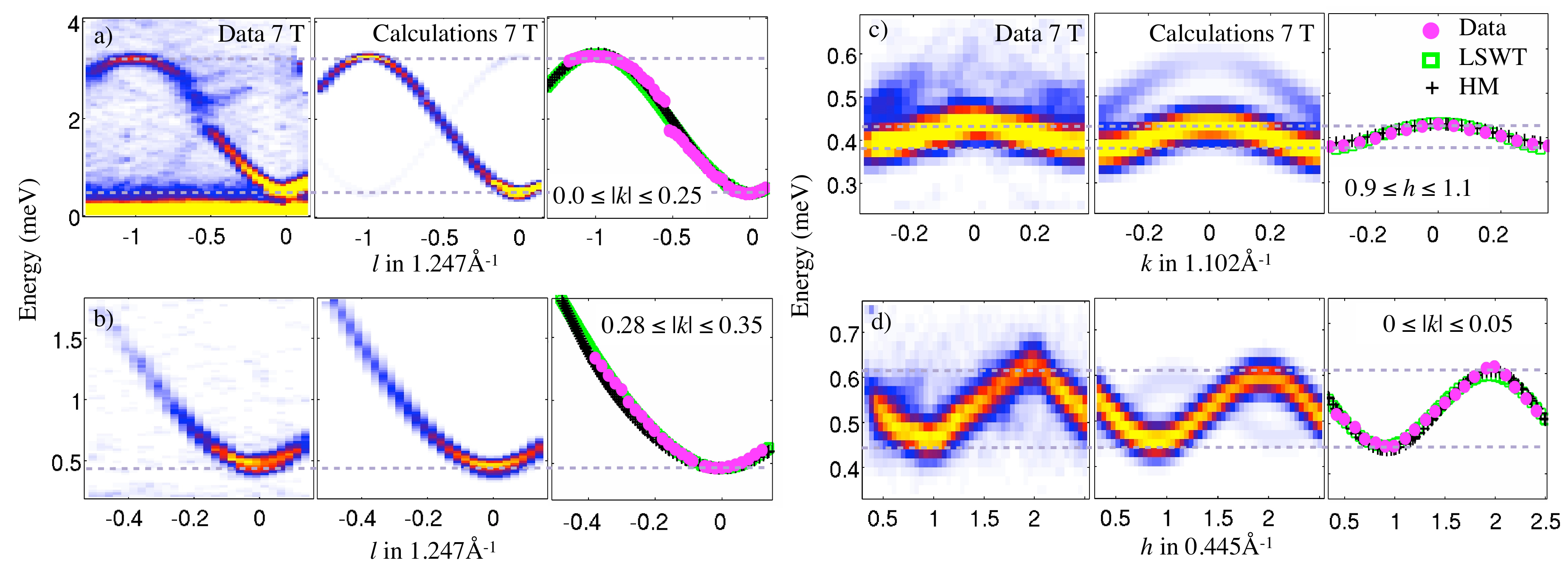}
\caption{\label{fig:HKLdisp} (Color online) Inelastic neutron
scattering intensity as a function of wave vector and energy
transfer measured on LET at 7\, T and $\sim$$0.06$\, K(left),
calculated intensity using the four-sublattice spin wave model
(center) described in Sec.\ \ref{sec:lswt_four} and comparison
between dispersion data points (magenta filled circles) with the
spin-wave dispersion relation Eq.~(\ref{hw}) (LSWT, green square)
and the hopping model Eq.~(\ref{hopping}) (HM, black cross)
(right). Horizontal gray dashed lines are guides to the eye to
emphasize the bandwidth of the dispersion along various
directions. The calculated intensities include the neutron
polarization factor, the spherical magnetic form factor for Co$^{2+}$,
and instrumental resolution effects. (a) Dispersion along the chain
direction $l$ showing the full bandwidth of 2.7\, meV
($E_i=10$\, meV). (b) Zoom into the low-energy part of the
dispersion along $l$ showing the energy gap at 0.48\, meV
($E_i=4$\, meV). (c) Dispersion along the interchain direction $k$
for wave vectors near $(1,k,0)$ with a bandwidth of 0.05\, meV. Note
in the data (left panel) the presence at large $|k|$ and energies
above the main mode of additional scattering intensity that
decreases rapidly as $k \rightarrow 0$. This is attributed (middle
panel) to the shadow mode that appears because of the unit cell
doubling in the $ab$ plane due to the alternate rotation of Ising
axes. (d) Dispersion along the interchain direction $h$ for
wave vectors near $(h,0.025,0)$ showing minima at odd $h$ and a
bandwidth of 0.16 meV.}

\includegraphics[width=\linewidth, clip=true,trim=-40 0 -40 0] {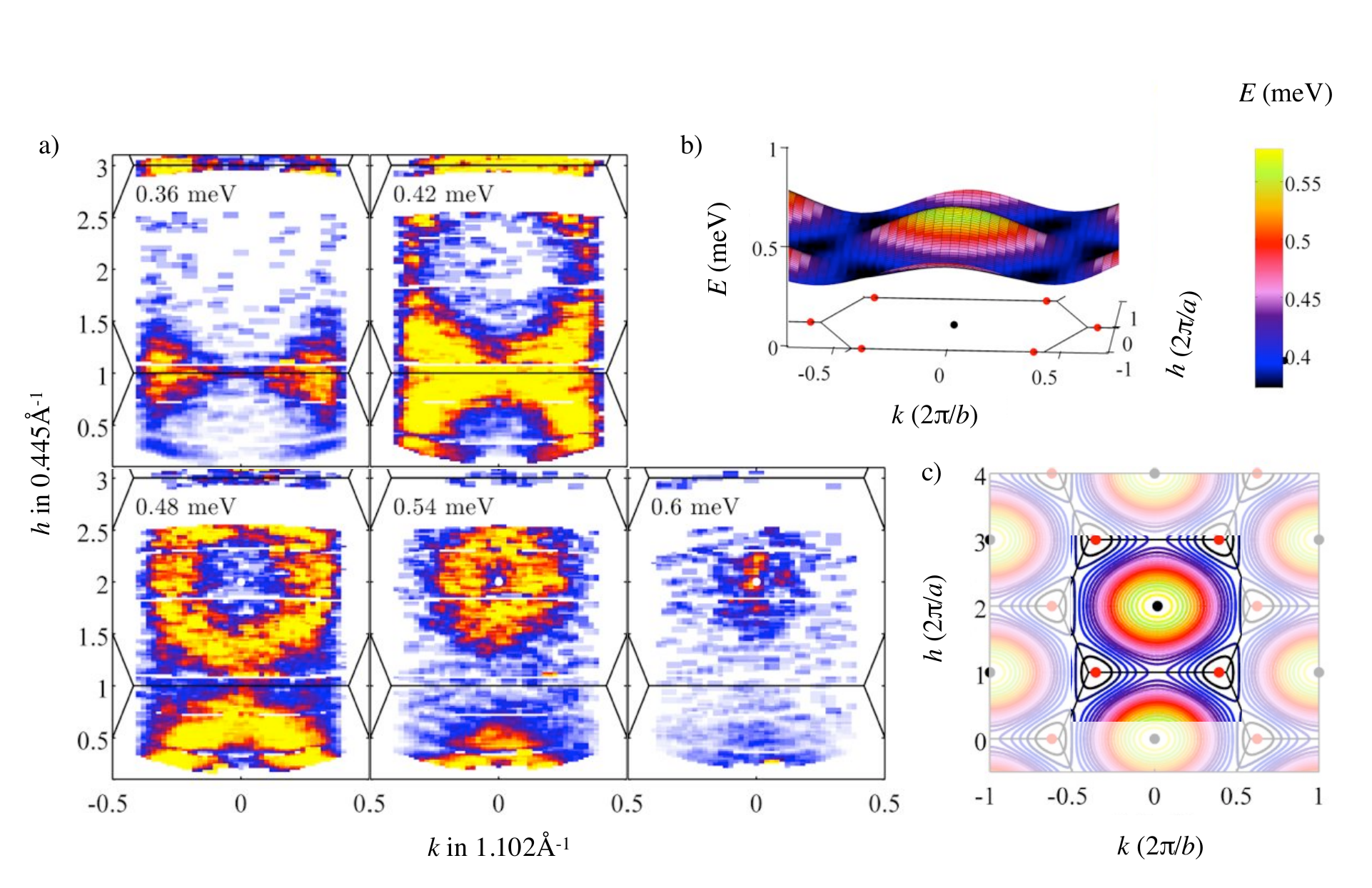}
\caption{\label{fig:letcontBZ} (Color online) (a) Observed neutron
scattering intensity in constant energy slices as a function of
wave vector in the $(hk0)$ plane at 7\, T and 1.8\, K, extracted from
a four-dimensional Horace scan on LET. Regions of strong intensity
show the constant-energy contours of the interchain dispersion in
the triangular $ab$ plane. Each slice shows data averaged for
energies within $\pm0.03$\, meV of the nominal value. Solid lines
show the edges of the hexagonal Brillouin zone of the triangular
lattice. (b) Surface plot of the interchain dispersion using
Eqs.~(\ref{hw}-\ref{fitted_exchanges}) and (c) projection contour
plot in the $(hk0)$ plane. Solid black circles show the Brillouin
zone centers (nuclear Bragg peak positions) of the triangular
lattice and solid red circles indicate minimum gap positions of
the dispersion surface. In (c) the highlighted area shows the
momentum space probed experimentally in (a).}
\end{figure*}
%%%%%%%%%%%%%%%%%%%%%%%%%%%%%%%%%%%%%

Higher-resolution measurements of the dispersion relations at this
same field (7\, T) illustrating the full bandwidth along the chain
direction, and along two orthogonal interchain directions are
shown in Fig.~\ref{fig:HKLdisp}(a)-(d). The dispersion bandwidth in
the $(hk0)$ plane is much smaller than the along-chain dispersion,
as expected for weakly-coupled chains (0.16\, meV and 0.05\, meV
along $(h00)$ and $(1k0)$, respectively, above a gap of 0.39\, meV,
compared to 2.7\, meV along $l$). The inelastic neutron scattering
intensity as a function of wave vector in the $(hk0)$ plane is
plotted for different energies in the panels of
Fig.~\ref{fig:letcontBZ}(a), where black lines denote the Brillouin
zone boundaries of the triangular lattice in the $ab$ plane. The
regions of strong scattering follow a dispersion resembling the
Fourier transform of a triangular lattice with antiferromagnetic
couplings showing a maximum energy near the zone centers (000) and
(200) (visible in the bottom right panel $E=0.6$\, meV) and minimum
energy near ($1,\pm q,0$) with $q\sim1/3$ (top left panel
$E=0.36$\, meV).

\section{Analysis}
\label{sec:analysis}
From the multi-dimensional inelastic neutron scattering data,
dispersion points were extracted by fitting Gaussian peaks to
scans in energy or wave vector to obtain the full
wave-vector $(h,k,l)$ and energy position $E$ of the intersection
of each scan direction with the dispersion surface. Since the LET
data provided high-resolution measurements of the dispersion
relations along all three directions in reciprocal space,
dispersion points ($h,k,l,E$) extracted from fitting datasets such
as those shown in magenta dots in Fig.~\ref{fig:HKLdisp} (a)-(d)
(right panels) were then used for the quantitative fits from the
models to be discussed below. The obtained parametrization was
checked for consistency against the MACS data for wave vectors in
the ($h0l$) plane.

\subsection{Spin-flip hopping model}

\label{sec:hopping} To parametrize the observed dispersions, we
first consider a phenomenological model of spin-flip excitations
that propagate by hopping between lattice sites, where the
dispersion is the Fourier transform of the hopping terms. We
assume a magnetic lattice of straight chains along $c$ ($\zeta=0$)
coupled in a triangular arrangement in the basal plane.  The
dispersion relation for nearest neighbor hops is
\begin{eqnarray}
\hbar\omega_{\bm{k}}&=&E_0+2t_0 \cos \pi l +\nonumber \\
 & & + 2t_1 \cos 2\pi k+4t_2 \cos\pi k ~ \cos \pi h. \label{hopping}
\end{eqnarray}
Here $E_0$ is the average energy (midpoint of the dispersion band)
and $t_0$, $t_1$ and $t_2$ describe the hoppings along the bonds
$\pm\bm{c}/2$ ($J_z$ bond), $\pm\bm{b}$ ($J_1$ bond) and
$\pm\frac{\bm{a}}{2}\pm\frac{\bm{b}}{2}$ ($J_2$ bond),
respectively, as illustrated in Fig.~\ref{fig:structure}. The
spin-flip hopping between sites physically originates from spin
exchange on the corresponding bonds with the hopping energy $t$
encoding the strength (and sign) of the spin interaction ($t>0$
for antiferromagnetic coupling). In the case of all bonds having
spin-isotropic exchanges of the type $J\bm{S}_{i}\cdot\bm{S}_{j}$,
the magnetic excitations in the fully-polarized phase at high
applied field are indeed single spin flips with hopping $t=SJ$;
for anisotropic exchanges and transverse fields, the dispersion is
expected to have a sinusoidal form as in Eq.~(\ref{hopping}) only
in the perturbative limit of small exchanges compared to the
Zeeman energy [\onlinecite{Sachdev}].

We find that at high field (7\, T) the overall shape and low-energy
modulations in the dispersion relation of the main mode can be
well described by the nearest-neighbor hopping model in
Eq.~(\ref{hopping}), see Fig.~\ref{fig:HKLdisp} (right panels,
magenta filled circles = data points and black crosses = model)
with fitted parameter values
\begin{equation}
\begin{array}{c c}
E_0= 1.857(3)\  \hbox{meV}, &  2t_0 = -1.402(2)\  \hbox{meV}\\
t_2 / t_1= 0.82(1), & 2t_1 = 0.0511(7)\  \hbox{meV}.\nonumber\\
\end{array}
\end{equation}
The model can reproduce the dispersion relation along the chain
direction ($t_0<0$ means ferromagnetic exchange along the chain),
see Fig.~\ref{fig:HKLdisp}(a)(right panel) as well as the
dispersions in the interchain ($hk0$) plane, which is that of a
triangular lattice with antiferromagnetic nearest-neighbor
couplings ($t_{1,2}>0$), see Fig.~\ref{fig:HKLdisp}(c)-(d)(right
panels). The fitted hopping parameters give $t_2/t_1<1$ as
expected for an isosceles triangular lattice.

\subsubsection{Shadow mode due to buckling of chains}

\begin{figure}[thb]
\includegraphics[width=\linewidth,clip=true,trim= -60 0 -60 0] {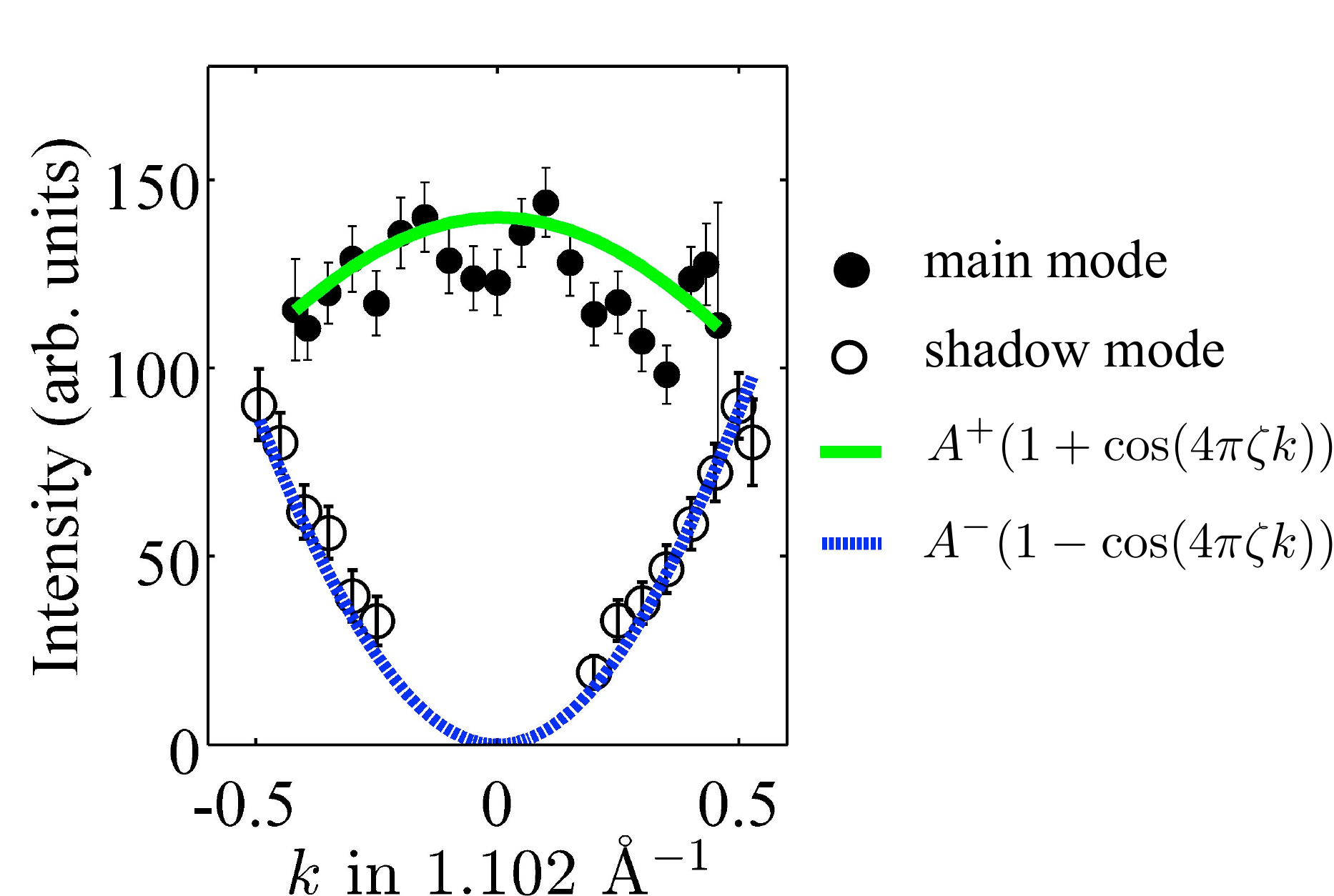}
\caption{\label{fig:kdep} (Color online) Intensity of the
main (shadow) mode [filled (open) circles] as a function of the
interchain wave vector $k$, fitted to the functional form for a
buckled magnetic chain [green solid (blue dashed) line], as
described in the text. The intensities were extracted at the
ferromagnetic zone boundary $l=-1$ from the same data set as in
Fig.~\ref{fig:Ldisp10meVall}. Error bars represent $\pm1$ standard
deviation.}
\end{figure}

The presence of the additional weaker intensity shadow mode in
Fig.~\ref{fig:Ldisp10meVall}(b) for finite $k$ can also be
explained within a hopping model by including the buckling of the
magnetic chains, see Fig.~\ref{fig:structure}(a), where consecutive
ions along the chain are alternatingly displaced by $\pm
\zeta\bm{b}$. This buckling along $b$ leads to a doubling of the
magnetic unit cell along $c$ (compared to straight chains), and
zone folding of the main mode dispersion, Eq.~(\ref{hopping}),
leads to the appearance of a second mode with the same dispersion
relation, but shifted in wave vector by $l \rightarrow l+1$, as
observed in Fig.~\ref{fig:Ldisp10meVall}(b). Using a spin-flip
hopping model on the corresponding two-sublattice problem, we
obtain the intensity of the two modes in inelastic neutron
scattering as $I^{\pm}_{\bm{k}}=A^{\pm} \left[ 1\pm \cos \left( 4
\pi \zeta k\right)\right]$, with the upper (lower) sign for the
main (shadow) mode. This predicts that the main mode is strongest
and the shadow mode absent at $k=0$, with the shadow mode intensity
increasing quadratically as $|k|$ increases from 0, which is
consistent with the observation of the shadow mode only at finite
$k$ in Fig.~\ref{fig:Ldisp10meVall}(a)-(b). Quantitatively, the
experimentally-observed $k$-dependence of the intensity of the two
modes extracted from the data shown in
Fig.~\ref{fig:Ldisp10meVall} is well described by the above
functional forms as shown in Fig.~\ref{fig:kdep} [data, open (filled)
circles; fits, solid (dashed) lines].

%%%%%%%%%%%%%%%%%%%%%%%%%%%%%%%%%%%%%%%%%%%%%%%
%%%%%%%%%%%%%%%%%%%%%%%%%%%%%%%%%%%%%%%%%%%%%%%%%%%%

We note however that the ratio of the extracted intensity
pre-factors of the shadow and main mode, $A^-/A^+=2.5(2)$ at
$l=-1$, cannot be accounted for within the hopping model, which
predicts $A^-=A^+$, both constants, independent of wave vector. The
hopping model predicts constant intensity for the dispersion along
$l$ (as expected for isotropic spin exchanges), whereas the data
clearly shows intensity decreasing upon increasing energy and
wave vector away from $l=0$ [see Fig.~\ref{fig:HKLdisp}(b) (left
panel)].  This intensity modulation with a maximum near $l=0$ (gap
minimum) is phenomenologically understood as an increase in
scattering weight at low energies in anticipation of the critical
phase transition at lower field near 5.5\, T.

Another shortcoming of the hopping model is that although it
accounts well for the observed modulations in the interchain
dispersion at the lowest energies near $l=0$, it predicts the same
magnitude interchain dispersion also at higher energies, whereas
experimentally it is observed that the dispersion relation becomes
more one-dimensional (less interchain dispersion) at higher
energies, as shown in Fig.~\ref{fig:macs0p40p9}, where constant
energy contours are strongly modulated along both $h$ and $l$ at
low energies, but they become almost independent of $h$ at high
energies above $\sim$1.25\, meV. Those shortcomings of the hopping
model in describing the intensity maximum near the gap minimum and
suppression of the interchain dispersion at higher energies are
better accounted for by including anisotropic spin exchange
interactions, as described in the following section.

\subsection{Linear spin-wave model}

\label{sec:lswt} To better account for the observed dispersion and
intensity modulations in the 3D Brillouin zone we parametrize the
data in terms of a microscopic spin Hamiltonian with anisotropic
spin exchanges.

\subsubsection{One-sublattice spin-wave model}

\label{sec:lswt_one} We start with straight chains along $c$
($\zeta=0$) with a dominant nearest-neighbor ferromagnetic Ising
exchange $J_z$ [see Fig.~\ref{fig:structure}] between the $S^z$
spin components and a smaller exchange $J_{xy}$ between the $S^x$
and $S^y$ spin components. As sub-leading terms, we include an
Ising second neighbor exchange $J'_z$ along the chains for the
$\pm\bm{c}$ bonds, and Ising interchain couplings $J_1$ and $J_2$
along the $\pm\bm{b}$ and $\pm\frac{\bm{a}}{2}\pm\frac{\bm{b}}{2}$
bonds, respectively \footnote{Although interactions between $S^x$ and $S^y$ spin components are allowed between further neighbors, we expect the resulting exchanges to be small enough to be neglected for the purposes of characterizing the dispersion relations in CoNb$_2$O$_6$.}. To start with, we also assume that the local
Ising axes ($z$) are the same for all sites [along the $c$ axis
($\gamma=0$)]. This ensures that we deal with a one-sublattice
magnetic unit cell of basis vectors $(\bm{a}-\bm{b})/2$, $\bm{b}$
and $\bm{c}/2$. Including also a magnetic field applied along $x$
(transverse to the Ising axis $z$), the Hamiltonian reads
\begin{eqnarray}
\label{Ham} \mathcal{H}=& \sum_{\bm{r}} -J_z S_{\bm r}^z
S_{\bm{r}+\bm{c}/2}^z
-J_{xy} \left[ S_{\bm r}^x S_{\bm{r}+\bm{c}/2}^x +S_{\bm{r}}^y S_{\bm{r}+\bm{c}/2}^y\right] \nonumber\\
 & + J'_z S_{\bm{r}}^z S_{\bm{r}+\bm{c}}^z + J_1 S_{\bm{r}}^z
 S_{\bm{r}+\bm{b}}^z+ J_2 S_{\bm{r}}^z S_{\bm{r}+(\bm{a}+\bm{b})/2}^z \nonumber\\
 & +J_2 S_{\bm{r}}^z S_{\bm{r}+(\bm{a}-\bm{b})/2}^z - g\mu_B
 BS^x_{\bm r},
\end{eqnarray}
where $g\mu_B B$ is the Zeeman energy, $S=1/2$, and $\bm{r}$ runs
over all magnetic lattice sites. We find that $J_z$ and $J_{xy}$
are ferromagnetic, whereas $J'_z$, $J_{1}$, and $J_2$ are
antiferromagnetic. The dispersion relations of the above
Hamiltonian are exactly solvable only in some special cases, such
as when $J_{xy}=J'_z=J_1=J_2=0$, i.e., decoupled Ising chains in a
transverse field [\onlinecite{Pfeuty}], and some related
models [\onlinecite{Kjall:2011fk}]. In the asymptotic limit of very high
fields when spins are nearly ferromagnetically polarized along the
applied field direction, the excitations can be described using
linear spin-wave theory. Our experiments were performed at applied
fields not high enough to be in this limit, but in the absence of
an alternative quantitative theory that could include all
exchanges, we have used linear spin-wave theory to parametrize the
excitations with the expectation that the fitted exchange values
might be renormalized from their actual values.

Assuming a mean-field fully-polarized ground state, the spin-wave
dispersion relation of the Hamiltonian in Eq.~(\ref{Ham}) is
obtained as
\begin{equation}
\hbar\omega_{\bm{k}}  = \sqrt{A_{\bm{k}}^2 - B_{\bm{k}}^2},
\label{hw}
\end{equation}
where
\begin{align}
A_{\bm{k}} & =  g\mu_B B -S[J_z(\bm{k})+J_{xy}(\bm{k})] +2SJ_{xy}(\bm{0}), \nonumber\\
B_{\bm{k}} & =  -S[J_z(\bm{k})-J_{xy}(\bm{k})], \label{AB}
\end{align}
and where the Fourier-transformed exchanges are
\begin{align}
\label{exchanges}
\begin{split}
J_z(\bm{k}) = &J_z\cos\pi l - J'_z\cos 2\pi l - J_1\cos 2\pi k\\
 & -2J_2\cos\pi k~\cos \pi h,\\
 J_{xy}(\bm{k}) =& J_{xy} \cos \pi l.
\end{split}
\end{align}
The intensity in neutron scattering is proportional to the
dynamical correlations obtained (see Appendix \ref{appendix}) as
$S^{zz}(\bm{k},E)=\frac{S}{2}\frac{A_{\bm{k}}-B_{\bm{k}}}{\hbar\omega_{\bm{k}}}~\delta\left(
E-\hbar\omega_{\bm{k}}\right)$ for the fluctuations polarized
along the Ising axis $z$, and
$S^{yy}(\bm{k},E)=\frac{S}{2}\frac{A_{\bm{k}}+B_{\bm{k}}}{\hbar\omega_{\bm{k}}}~\delta\left(
E-\hbar\omega_{\bm{k}}\right)$ for the fluctuations along $y$, the
axis perpendicular to the plane defined by the Ising axis and the
applied field. Here $\delta$ is the Dirac delta function. For
direct comparison with the experiment, the expected neutron
scattering intensity including polarization factors is given by
\begin{equation}\label{intensities}
I(\bm{k},E)=
\left({1-\frac{\kappa_y^2}{\kappa^2}}\right)S^{yy}(\bm{k},E)+
\left({1-\frac{\kappa_z^2}{\kappa^2}}\right)S^{zz}(\bm{k},E)
\end{equation}
where
\begin{equation}
\kappa_y = 2\pi h/a, \ \ \ \ \kappa_z= 2\pi l/c, \ \ \ \
\kappa=\sqrt{\kappa_y^2+\kappa_z^2}.
\end{equation}
The strongest intensity is predicted for fluctuations polarized
along the Ising axis $z$ at low energies near the 3D dispersion
minima points at (odd, $q$, 0) and (even, $1-q$, 0) with
$q=\frac{1}{\pi}\cos^{-1}\left(\frac{J_{2}}{2J_1}\right)$. A
contour plot of the dispersion surface in the $(hk0)$ plane is
shown in Fig.~\ref{fig:letcontBZ}(c).

\subsubsection{Four-sublattice spin-wave model}
\label{sec:lswt_four}

The one-sublattice spin Hamiltonian in Eq.~(\ref{Ham}) can be
readily extended to the full generality of the actual crystal
structure, where the magnetic chains along $c$ are not straight,
but buckled [see Fig.~\ref{fig:structure}(a)], and the local Ising
axes are not the same for all sites, but alternate in orientation
between $\hat{\bm{z}},\hat{\bm{z}'}=\pm\sin\gamma \hat{\bm{a}} +
\cos \gamma \hat{\bm{c}}$ for magnetic chains translated by
$(\pm\bm{a}\pm\bm{b})/2$. For example, if we were to identify the
local Ising axes in Fig.~\ref{fig:structure}(b), even and odd
$b$ axis ``rows'' of spins would show that the local Ising axes
alternate between $\hat{\bm{z}}$ and $\hat{\bm{z}'}$. Here
$\hat{\bm{a}}$ and $\hat{\bm{c}}$ are unit vectors along the
orthorhombic $a$ and $c$ axes, respectively. To compare the
results directly with the one-sublattice model, we will assume that
the interchain exchanges still couple only the local Ising spin
components on the different chains, i.e., $J_1S^z_{\bm{r}}S^{z'}_{\bm{r}+\bm{b}}$ and $J_2S^z_{\bm
r}S^{z'}_{\bm{r}+(\bm{a}\pm\bm{b})/2}$.

By solving this four-sublattice model (the magnetic unit cell
equals the structural, orthorhombic unit cell) via numerical
diagonalization of the quadratic spin-wave Hamiltonian, we
obtained four dispersive modes, one mode has the same dispersion,
Eq.~(\ref{hw}), as the one-sublattice problem in Eq.~(\ref{Ham}),
and the other three modes are obtained by a shift in wave vector.
This can be intuitively understood starting from the
one-sublattice problem: the buckling of the chains and
non-equivalence of Ising axes between chains translated by
$(\pm\bm{a}\pm\bm{b})/2$ leads to a doubling of the magnetic unit
cell along $c$ and also a doubling in the $ab$ plane. New magnetic
zone centers appear at positions such as (001) and (100), and
consequently new dispersion modes appear, which are images of the
main mode shifted in wave vector to the new zone center positions.
In total, three additional shadow modes appear with dispersion
relations $\hbar\omega_{(h,k,l+1)}$ [as seen in
Fig.~\ref{fig:Ldisp10meVall}(b)], $\hbar\omega_{(h+1,k,l)}$ [as
seen in Fig.~\ref{fig:HKLdisp}(c) (left and middle panels)], and
$\hbar\omega_{(h+1,k,l+1)}$.

The experimentally-observed dispersion relation of the main mode
can be well-described by the spin-wave model in Eq.~(\ref{hw}),
when all exchange values are included and have finite values
(attempts to fit the dispersion to a restricted set of parameters,
such as fixing $J_{xy}=0$ or $J_z'=0$ did not provide good fits).
However, the dispersion data alone are not sufficiently
constraining to independently determine the absolute values of all
five exchanges and the Zeeman term because changes in parameter
values produce changes to the dispersion that are strongly coupled
to one another. In particular, the effects of $J_z$ and $J_{xy}$
are to vary the dispersion bandwidth by affecting both the minimum
and maximum energies; the $B$-field overall shifts the dispersion
to higher energies, but it also affects slightly the dispersion
bandwidth; $J'_z$ overall shifts the dispersion and produces a
modulation with periodicity $l\rightarrow l+1$. Even the
dispersion bandwidth in the plane normal to the chains depends not
only on the strength of the interchain couplings $J_{1,2}$, but
also the Zeeman term $g\mu_BB$, as can be explicitly seen by
inspecting the form of the dispersion relation at $l=0$,
$\hbar\omega_{(h,k,0)}=$
$$\sqrt{C+2g\mu_BBS\left[J_1\cos 2\pi k +
2J_2\cos \pi k ~ \cos \pi h\right]},$$ where
$C=g\mu_BB\left[g\mu_BB-2S\left(J_z-J_z'-J_{xy}\right)\right]$. Note that
the pre-factor in front of the Fourier transfom of the
interchain terms is not a constant, but depends on the Zeeman
term. So effectively the absolute values of {\em all} exchange
parameters and the Zeeman term are strongly coupled to one another
within the spin-wave dispersion form, and as a consequence several
sets of parameter values can give rather similar agreement with
the observed dispersions, as we indeed find. The only parameter
that is relatively constrained by the parametrizations (and for which we can provide a meaningful uncertainty) is the
interchain frustration ratio $J_2/J_1$. To illustrate the level of
agreement that can be obtained, we list below {\em representative}
values for the parameters that give one of the lowest $\chi^2$
values in terms of a comparison with the dispersions at 7\, T
\begin{equation}
\begin{array}{c c}
J_z= 2.19\  \hbox{meV} &  g\mu_B B = 1.66\  \hbox{meV}\\
J_{xy} = 0.36\  \hbox{meV} &  J'_z = 0.29\  \hbox{meV}\\
J_1=0.031 \  \hbox{meV} & J_{2}/J_1 = 0.77(10)\  \hbox{meV}
\label{fitted_exchanges}
\end{array}
\end{equation}
and at 9\, T
\begin{equation}
\begin{array}{c c}
J_z= 2.4\  \hbox{meV} &  g\mu_B B = 2.0\  \hbox{meV}\\
J_{xy} = 0.4\  \hbox{meV} &  J'_z = 0.36\  \hbox{meV}\\
J_1 = 0.036 \  \hbox{meV} & J_2/J_1= 0.75(10)\ \hbox{meV}.
\label{fitted_exchanges9T}
\end{array}
\end{equation}
These parametrizations capture well all key modulations of the
observed dispersions. They also reproduce well the overall
intensity dependence (using the four-sublattice model), as
illustrated in Figs.~\ref{fig:macs0p40p9}, \ref{fig:HKLdisp}, and
\ref{fig:letcontBZ}. Furthermore, they reproduce quantitatively
the observed relative intensity ratio between the main and shadow
modes in Fig.~\ref{fig:kdep}, and they capture the fact that the
observed interchain dispersion is largest at the lowest energies
near $l=0$ and becomes progressively less pronounced at higher
energies.

We note that in all parametrizations, the ratio of the interchain
exchange couplings is consistent with the expectation of an
isosceles triangular lattice ($J_2/J_1 < 1$). This would give an
instability to magnetic order at an incommensurate wave vector
$q=\frac{1}{\pi}\cos^{-1}\left(\frac{J_2}{2J_1}\right)\sim0.37(2)$,
consistent with the value observed experimentally at the onset of
magnetic order, $q(T_{\rm N}=2.95~{\rm K})=0.37$ (see
Ref.~[\onlinecite{Heid1995123}]), further supporting the idea that
frustrated interchain couplings stabilize the incommensurate spin-density-wave magnetic order observed at zero field just below
$T_{\rm N}$.

For completeness, we note that an unexpected feature in the data,
not captured by the spin-wave model, is an apparent broadening of
the magnetic scattering intensity and a departure of the local
dispersion slope away from the model predictions in a finite
energy range just above the midpoint of the dispersion bandwidth
along the chain direction; see the 7\, T data in
Fig.~\ref{fig:HKLdisp}(a)(left panel) in the approximate energy
range $1.8-2.3$\, meV. Similar anomalies are also observed upon
close inspection in the 9\, T data in
Fig.~\ref{fig:Ldisp10meVall}(a) in a finite energy range
($2.4-3$~meV), again just above the midpoint of the dispersion
bandwidth. Those anomalous broadening effects at intermediate
energies will be addressed in detail elsewhere [\onlinecite{Robinson}].

\section{Discussion}
\label{sec:discussion} We now compare the parametrization of the
dispersions with previous estimates of the exchange couplings
obtained from analyzing the excitations in zero
field [\onlinecite{Coldea}], where the spectrum consisted of a series of
sharp modes strongly dispersing along the chain direction and
attributed to bound states of pairs of domain walls (kinks) on the
magnetic chains. This spectrum was well explained by an effective
Hamiltonian for kinks [\onlinecite{Coldea,Ruthkevich}], which contained the
energy cost $J$ required to create two kinks in the absence of
other perturbations, a kink hopping term, $\alpha$, tuning the
dispersion bandwidth, and terms $\beta$ and $\beta'$ to account
for the energy of a kinetic bound state stabilized near the
ferromagnetic zone boundary ($l=-1$). In addition, a longitudinal
effective field $h_z$ was assumed responsible for the confinement
of pairs of kinks into bound states. Mapping the spin Hamiltonian
(\ref{Ham}) at zero external field ($B=0$) into an effective
Hamiltonian for kinks in the limit of small perturbations from the
Ising limit ($J_z$ dominant) reproduces several terms in the
phenomenological kink Hamiltonian in Ref.\ [\onlinecite{Coldea}].  These
terms are $J=J_z-2J'_z$, $\beta =J_{xy}/2$, $\beta'=J_z'$ and the
mean field due to 3D long-range order in the antiferromagnetic
pattern with $q_{\rm AF}=(0,1/2,0)$ is $h_z=2 J_1 \langle S^z
\rangle$, where $\langle S^z \rangle$ is the expectation value of
the ordered spin moment.  The kink hopping term $\alpha$ is not
captured in this mapping; it must originate microscopically from a
magnetic interaction term not considered in Eq.~(\ref{Ham}). Using
the zero-field values for $J$, $\beta$, and $\beta'$ from Ref.\ [\onlinecite{Coldea}] gives $J_z=2.76$\, meV, $J_{xy} = 0.66$\, meV,
and $J'_z = 0.41$\, meV. We note that a more elaborate analysis of
the zero-field excitation spectrum (using numerical matrix-product
state methods for 1D chains expected to be accurate even in the
presence of substantial perturbations away from the Ising limit)
proposed somewhat similar values [\onlinecite{Kjall:2011fk}],
$J_z=2.43$\, meV, $J_{xy} =0.52$\, meV, and $J'_z =0.60$\, meV. The
observed dispersion at 7 and 9 \, T cannot be quantitatively
described by either of the above two sets of exchanges by only
allowing as free parameters the Zeeman energy and the interchain
couplings $J_1$ and $J_2$; the observed dispersion bandwidth along
the chain direction is systematically smaller than the prediction.
We attribute this effect to a quantum renormalization on the
dispersion relation in the quantum paramagnetic phase beyond the
linear spin-wave approximation, as explained below.

\setcounter{subsubsection}{0}
\subsubsection{Quantum renormalization effects on the spin-wave dispersion}

To appreciate why spin-wave theory does not capture quantitatively
the dispersion relation in the quantum paramagnetic phase, it is
insightful to consider the pure 1D Ising chain in transverse
field, i.e.,
\begin{equation}
{\cal H}=\sum_i-J_zS^z_iS^z_{i+1}-hS^x_i,
\end{equation}
where $i$ indexes consecutive sites along the chain, $J_z$ is the
ferromagnetic Ising coupling, $h$ is the transverse field and
$S=1/2$. The semi-classical, mean-field approach predicts
suppression of the spontaneous ferromagnetic Ising order at the
classical critical field $h_{c}^{cl}=J_z$ with the spin-flip
dispersion at higher field
\begin{equation}
\hbar\omega^{cl}=\sqrt{h^2-hJ_z\cos \pi l},
\label{classical_dispersion}
\end{equation}
(for spacing $c/2$ along the chain). In contrast, the exact
quantum solution (obtained via mapping to Jordan-Wigner
fermions [\onlinecite{Pfeuty,Sachdev}]) gives the critical field at {\em
half the classical value}, i.e., $h_c=J_z/2$, with the
quasiparticle dispersion at higher field
\begin{equation}
\hbar\omega=\sqrt{h^2-hJ_z\cos \pi l+\frac{J^2_z}{4}}.
\label{quantum_dispersion}
\end{equation}
Both the classical and quantum dispersions tend to the same form,
$h-\frac{J_z}{2} \cos \pi l$, in the perturbative limit near very
high field $h/J_z\rightarrow \infty$, but there are very
significant differences at fields comparable to the exchange
strength. The strong renormalization of the critical field means
that for fields in the range $J_z/2 < h \lesssim J_z$ the
spin-wave description, which assumes a polarized ground state,
would be unstable.

The classical and quantum dispersions at their respective critical
fields have the same functional form (sinusoidal), but predict
different dispersion bandwidths, i.e., $\hbar \omega (h=h_c)= J_z
\left| \sin \frac{\pi l}{2}\right|$ compared to
$\hbar\omega^{cl}(h=h^{cl}_c)= \sqrt{2}J_z \left| \sin \frac{\pi
l}{2}\right|$, so the spin-wave formula
Eq.~(\ref{classical_dispersion}) can be used to ``fit'' the
quantum dispersion at the actual critical field $h_c$, but using a
renormalized exchange $\tilde{J_z}=J_z/\sqrt{2}$ and a
renormalized Zeeman energy $\tilde{h}=\sqrt{2}h$, i.e., the
``fitted'' exchange would appear $\sim$$30\%$ smaller than the
actual value and one would have to use an artificially-larger
Zeeman term. For fields above $h_c$ the classical and quantum
dispersions do not have the same functional form, but one can
approximately ``fit'' the quantum dispersion
(\ref{quantum_dispersion}) with a classical relation
(\ref{classical_dispersion}) with renormalization factors for
$J_z$ and $h$ that progressively tend to unity in the limit of
very high field $h/J_z\rightarrow\infty$.

Based on the above discussion, we propose that quantum
renormalizations of the dispersion not captured by a spin-wave
approach in the region of transverse fields slightly above the
critical field are responsible for the apparently smaller
dispersion bandwidth than predicted. We note that the empirically
extracted renormalization at 9\, T is smaller compared to 7\, T.
This is consistent with the expectation that quantum
renormalization of the bandwidth decreases upon increasing field
closer to the high-field limit. Future experiments at sufficiently
large fields could provide a test for where renormalization
effects become negligible and the classical and quantum
descriptions become equivalent.
\\

\section{Conclusions}
\label{sec:conclusions} We have reported a comprehensive study of
the magnetic dispersion relations using inelastic neutron
scattering in the quasi-1D Ising ferromagnet CoNb$_2$O$_6$ in the
quantum paramagnetic phase in a high transverse magnetic field. The
spectrum is dominated by a sharp mode, as expected for
coherently-propagating spin-flip quasiparticles. In addition to
the main dispersive mode, much weaker intensity shadow modes were
also observed and attributed to the enlargement of the magnetic
unit cell due to the buckling of the magnetic chains and the
alternating rotation of Ising axes between chains. The largest
dispersion is observed along the chain direction $l$, with clear
modulations in the dispersion along $h$ and $k$ at the lowest
energies due to the interchain couplings, which form an isosceles
triangular lattice geometry. The observed dispersions have been
parameterized by a phenomenological spin-flip hopping model and
also by a linear spin-wave model. Differences in the observed
dispersion bandwidth along the chain direction and spin-wave
prediction using estimated exchange values from analysis of
zero-field dispersions are attributed to strong quantum
renormalization effects of the dispersion relation in the quantum
paramagnetic phase not captured by a linear spin-wave approach for
fields slightly above the critical field, where quantum
fluctuations in the ground state are still significant.

\section{Acknowledgements}
We acknowledge very useful discussions with F. H. L. Essler and N.
J. Robinson. IC and RC acknowledge support from the EPSRC Grant
No. EP/H014934/1. JDT was supported by the University of Oxford
Clarendon Fund Scholarship and NSERC of Canada. This work utilized
facilities supported in part by the National Science Foundation
under Agreement No. DMR-0944772. We acknowledge STFC (UK) and the
National Institute of Standards and Technology, U.S. Department of
Commerce in providing the neutron research facilities used in this
work and we thank the technical staff at those facilities for
cryogenics support. We acknowledge collaboration with E. M. Wheeler
on preliminary measurements of magnetic excitations at lower
applied magnetic fields at the Helmholtz-Zentrum
Berlin [\onlinecite{Wheeler}].

%%%%%%%%%%%%%%%%%%%%%%%%%%%%%%%%%%%%%%%%%%%%%%%%%%%%%%%%%%%%%%%%%%%%%
\appendix
\section{Dynamical correlations in the spin-wave model}
\label{appendix}
%%%%%%%%%%%%%%%%%%%%%%%%%%%%%%%%%%%%%%%%%%%%%%%%%%%%%%%%%%%%%%%%%%%%%
For completeness we outline here the derivation of the dynamical
structure factor for the spin-wave model for the Hamiltonian in
Eq.\ (\ref{Ham}) in the the limit of high magnetic fields $B$,
where the mean-field ground state has fully-polarized spins along
the field direction. It is useful to relabel the axes as
$(x,y,z)=(z',-y',x')$ such that the equilibrium spin direction is
along $z'$ (and the Ising axis is along $x'$). Using a
Holstein-Primakoff transformation to spin deviation operators (for
each spin site $S^{z'}=S-a^{\dag}a, S^{+}\simeq\sqrt{2S}a,
S^{-}\simeq\sqrt{2S}a^{\dag}$), the leading form of the Hamiltonian
expansion in terms of spin-wave operators is
$\mathcal{H}=NE_{\rm{MF}}(1+1/S)+\sum_{\bm k}
\mathsf{X}^{\dagger}\mathsf{H}\mathsf{X}$. The first term contains
$E_{\rm{MF}}$, the mean-field ground state energy normalized to
the number of sites ($N$), and the second term is a quadratic form
of Bose operators written in a matrix form using the operator
basis $\mathsf{X}^{\dagger}=[a_{\bm k}^{\dag} ~, ~ a_{-\bm k}]$
and the Hamiltonian matrix
\begin{equation}
\mathsf{H}=\frac{1}{2}\left[\begin{array}{ll}
A_{\bm k} &  B_{\bm k}\\
B_{\bm k} &  A_{\bm k}
\end{array}\right].
\end{equation}
Here $a_{\bm k}^{\dag} =\frac{1}{\sqrt{N}}\sum_{\bm r} e^{-i{\bm
r}\cdot {\bm k}} a_{\bm r}^{\dag} $ creates a spin wave of
momentum ${\bm k}$ (the sum extends over all spin sites ${\bm
r}$). Expressions for $A_{\bm k}$ and $B_{\bm k}$ (both real) are
given in Eq.\ (\ref{AB}). Diagonalizing the quadratic Hamiltonian
form using standard methods [\onlinecite{White}] gives the dispersion
relation $\hbar\omega_{\bm k}=\sqrt{A_{\bm k}^2-B_{\bm k}^2}$ and
the transformation between the original basis and the normal
magnon basis $\mathsf{X'}^{\dagger}=[\alpha_{\bm k}^{\dag} ~, ~
\alpha_{-\bm k}]$, $\mathsf{X}=\mathsf{S}\mathsf{X'}$, is obtained
as
\begin{equation}
\mathsf{S}=\left[\begin{array}{rr}
u_{\bm k} &  -v_{\bm k}\\
-v_{\bm k} &  u_{\bm k}
\end{array}\right],
\end{equation}
where $u_{\bm k}=\cosh \theta_{\bm k}$, $v_{\bm k}=\sinh
\theta_{\bm k}$, and $\tanh 2\theta_{\bm k}=B_{\bm k}/A_{\bm k}$.
In the normal magnon basis, the Fourier-transformed spin operator
along the Ising axis is $S^{x'}_{\bm
k}=\sqrt{\frac{S}{2}}\left(u_{\bm k}-v_{\bm
k}\right)\left[\alpha_{\bm k} + \alpha_{-\bm{k}}^{\dag}\right]$.
From this, the dynamical correlations along the $x'$ axis are
obtained as $S^{x'x'}({\bm k},E)=\frac{S}{2}|u_{\bm k}-v_{\bm
k}|^2~\delta(E-\hbar\omega_{\bm k})=\frac{S}{2}\frac{A_{\bm
k}-B_{\bm k}}{\hbar\omega_{\bm k}}~\delta(E-\hbar\omega_{\bm k})$.
Similarly, $S^{y'y'}({\bm k},E)=\frac{S}{2}|u_{\bm k}+v_{\bm
k}|^2~\delta(E-\hbar\omega_{\bm k})=\frac{S}{2}\frac{A_{\bm
k}+B_{\bm k}}{\hbar\omega_{\bm k}}~\delta(E-\hbar\omega_{\bm k})$,
giving the expressions listed in the text after Eq.\
(\ref{intensities}).

\bibliography{ConboICDpaper}
\end{document}